\documentclass[preprint,flushrt]{aastex}

\usepackage[utf8]{inputenc}
\usepackage{amssymb}
\usepackage{amsmath}
\usepackage{graphicx}
\usepackage{color}
\usepackage{natbib}
\usepackage{booktabs}
\bibpunct{(}{)}{;}{a}{}{,}

\newcommand{\mat}[1]{\boldsymbol{#1}}
\renewcommand{\vec}[1]{\boldsymbol{#1}}

\shorttitle{Parallax of the \emph{IBEX} Ribbon}
\shortauthors{Swaczyna et al.}
\slugcomment{}
 
\begin{document}

\title{Distance to the \emph{IBEX} Ribbon Source Inferred from Parallax}

\author{
P.~Swaczyna\altaffilmark{1},
M.~Bzowski\altaffilmark{1},
E.~R.~Christian\altaffilmark{2},
H.~O.~Funsten\altaffilmark{3},
D.~J.~McComas\altaffilmark{4,5},
N.~A.~Schwadron\altaffilmark{4,6}
}
\email{pswaczyna@cbk.waw.pl}

\altaffiltext{1}{Space Research Centre of the Polish Academy of Sciences (CBK PAN), Bartycka 18A, 00-716 Warsaw, Poland}
\altaffiltext{2}{Goddard Space Flight Center, Greenbelt, MD 20771, USA}
\altaffiltext{3}{Los Alamos National Laboratory, Los Alamos, NM 87545, USA}
\altaffiltext{4}{Southwest Research Institute, San Antonio, TX 78228, USA}
\altaffiltext{5}{University of Texas at San Antonio, San Antonio, TX 78249, USA}
\altaffiltext{6}{University of New Hampshire, Durham, NH 03824, USA}

\begin{abstract}
Maps of Energetic Neutral Atom (ENA) fluxes obtained from \emph{Interstellar Boundary Explorer} (\emph{IBEX}) observations revealed a bright structure extending over the sky, subsequently dubbed the \emph{IBEX} ribbon. The ribbon had not been expected from the existing models and theories prior to \emph{IBEX}, and a number of mechanisms have since been proposed to explain the observations. In these mechanisms, the observed ENAs emerge from source plasmas located at different distances from the Sun. Since each part of the sky is observed by \emph{IBEX} twice during the year from opposite sides of the Sun, the apparent position of the ribbon as observed in the sky is shifted due to parallax. To determine the ribbon parallax, we found the precise location of the maximum signal of the ribbon observed in each orbital arc. The obtained apparent positions were subsequently corrected for the Compton-Getting effect, gravitational deflection, and radiation pressure. Finally, we selected a part of the ribbon where its position is similar between the \emph{IBEX} energy passbands. We compared the apparent positions obtained from the viewing locations on the opposite sides of the Sun, and found that they are shifted by a parallax angle of $0.41^\circ\pm0.15^\circ$, which corresponds to a distance of $140^{+84}_{-38}$~AU. This finding supports models of the ribbon with the source located just outside the heliopause. 
\end{abstract}

\section{INTRODUCTION}
\label{sec:introduction}
Energetic Neutral Atoms (ENAs) are an important source of information about the interaction of the solar wind with the plasma of the local interstellar medium (LISM) \citep{gruntman_etal:01a}. ENAs are created from energetic ions via charge exchange with background neutral atoms. This process preserves the instantaneous velocity of the ions, and the created ENAs are no longer subject to electromagnetic forces. The ENA trajectories are only deflected by gravity and radiation pressure. The \emph{Interstellar Boundary Explorer} \citep[\emph{IBEX},][]{mccomas_etal:09a} observes ENA fluxes at Earth's orbit. First results of these observations revealed an arc-like enhancement in the flux on the sky \citep{mccomas_etal:09c}, dubbed the \emph{IBEX} ribbon. The ribbon had not been expected from models of the heliosphere--LISM interaction available before the \emph{IBEX} launch \citep{schwadron_etal:09b}. 

\citet{mccomas_etal:09c} proposed several alternative mechanisms of the ribbon generation, while subsequent authors proposed further models or refined the existing ones. For comprehensive review of hypothesized models see \citet{mccomas_etal:14a}. One of the consequences of these different models are different distances from the Sun to the regions where the ribbon ENAs are generated. The source closest to the Sun was proposed by \citet{kucharek_etal:13a}, where the ENAs are created just beyond the termination shock. Other models propose ribbon emission in the inner heliosheath \citep{mccomas_etal:09c, fahr_etal:11a, fichtner_etal:14a, sylla_fichtner:15}, at the heliopause \citep{mccomas_etal:09c}, and in the outer heliosheath \citep{mccomas_etal:09c, heerikhuisen_etal:10a, chalov_etal:10a, schwadron_mccomas:13a}. In all of these models, the ENAs emerge from processes resulting from the interaction of the solar wind and the LISM. A mechanism of the ENA generation beyond the heliosphere at the interface between the Local Interstellar Cloud with a bay in the Local Bubble was proposed by \citet{grzedzielski_etal:10b}. In this scenario, the distance to the ribbon source is likely the largest ($\gtrsim  300$~AU).  

One of the models proposed by \citet{mccomas_etal:09c} is the secondary ENA model that involves a heliospheric plasma ion that undergoes three sequential charge exchange events before detection by \emph{IBEX}. In this model, ENAs produced in the heliosphere are reionized in the outer heliosheath, and after subsequent neutralization via a final charge-exchange process, they produce secondary ENAs, observed by \emph{IBEX}. Several variants of the scenario were proposed. \citet{mccomas_etal:09c} were the first to suggest that neutrals from the solar wind may form pickup ion ring distributions. Reneutralized neutral particles from the pickup ion ring may move back through the heliosphere and be observed by \emph{IBEX} provided that the plane of the pickup ring is in the line-of-sight of the \emph{IBEX} ENA sensors. This implies that the \emph{IBEX} ribbon is observed only where the line-of-sight is roughly perpendicular to the interstellar magnetic field. The process was subsequently modeled using magnetohydrodynamic simulations \citep{heerikhuisen_etal:10a, heerikhuisen_pogorelov:11a, pogorelov_etal:11a, zirnstein_etal:15a} and using analytical theory to understand its properties \citep{mobius_etal:13a}. \citet{schwadron_mccomas:13a} propose a variant of this scenario in which the ribbon was modeled as a spatial "retention region" where pickup ions are scatttered by wave-particle interactions. Several others models of spatial retention have also been proposed by \citet{isenberg:14a} and \citet{giacalone_jokipii:15a}.

Discrimination between these different models would be an important milestone for understanding the interaction between the solar wind and the LISM. \citet{mccomas_etal:12b} argued that the energy-dependent structure of the ribbon should result from the helio-latitudinal structure of the solar wind over the prior several years \citep{sokol_etal:13a,sokol_etal:15d}. \citet{schwadron_mccomas:13a} were the first to incorporate detailed latitude dependence of the solar wind into a secondary ribbon model (the spatial retention model) and demonstrate a remarkable consistency with observed ribbon ENAs. In the secondary ENA model, the effect of solar cycle should be reflected in the ribbon fluxes with a temporal delay dependent on the distance to the source. \citet{zirnstein_etal:15b} simulated this effect using time-dependent solar wind boundary conditions. \citet{swaczyna_etal:14a} proposed that an attempt to observe the ribbon in helium ENAs should allow discrimination between the secondary ENA model and the extraheliospheric source model proposed by \citet{grzedzielski_etal:10b}.

The ribbon scribes a nearly complete circular arc that extends over a large part of the sky \citep{funsten_etal:09b}. The ribbon is a sharp ENA flux peak that is observed in all \emph{IBEX}-Hi energy passbands \citep{fuselier_etal:09b}. Its ENA intensity varies significantly around the ribbon and across the energy passbands. Because of its strong circularity, the location of the ribbon as projected in the sky is defined using a ribbon center at ecliptic coordinates $(\lambda_\text{RC},\beta_\text{RC})=(219.2^\circ,\,39.9^\circ)$ and a half-cone angle $\alpha=74.5^\circ$ that are derived using the average ribbon location at several energy passbands. The location of the ribbon is observed to vary slightly as a function of energy.
\citet{funsten_etal:15a} found that the ribbon features a symmetry that changes from unimodal to bilateral with increasing energy. 

In this paper, we determine the geometrical parallax of the ribbon and obtain the distance from the Sun to the plasma source. This allows us to discriminate between different models of the ribbon.

\section{PARALLAX OF ENA SOURCES}
\label{sec:trajectories}
Observation of the parallax of a celestial body is a time-tested technique in astrometry. Determining distances to the stars by measurement of their parallaxes is model independent and thus is used to scale other astronomical methods of distance determination. In astrometry, parallax is used for point sources (stars), or for bodies with well defined geometrical centers (e.g., planets). In contrast, the \emph{IBEX} ribbon is not a point source and is dispersed over a large region of the sky. Nevertheless, the ribbon is a strongly-peaked flux distribution. Using the measured ENA flux we identify positions of the flux maximum along directions that intersect with the ribbon. The set of the positions of the flux maxima yields a curve on the sky which we refer to as the ribbon ridge. The apparent position of the ridge observed from the opposite sides of the Sun is shifted relatively to the heliocentric location due to parallax.

\begin{figure}[ht!]
\epsscale{.5}
 \plotone{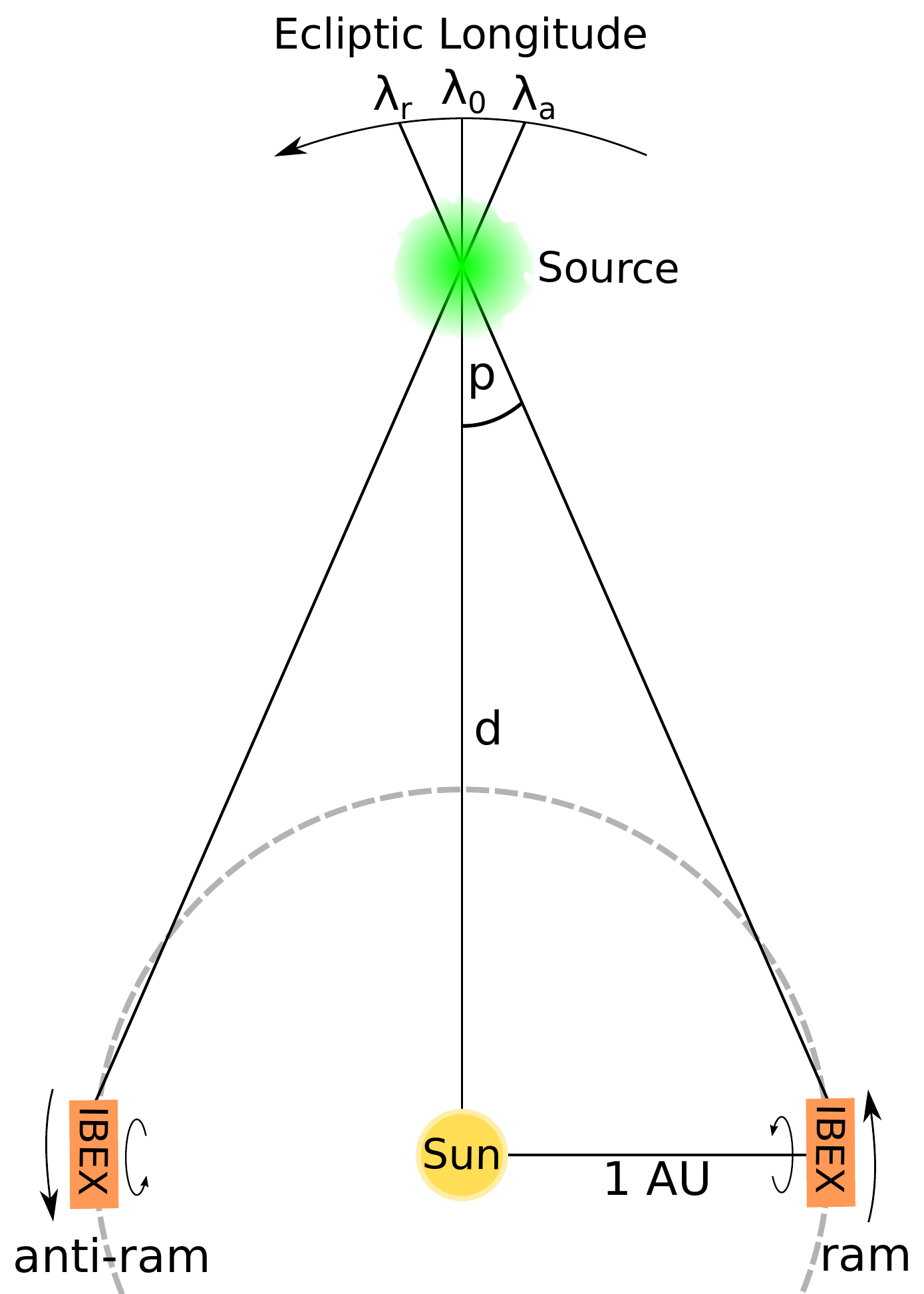}
 \caption{A illustration of the parallax geometry of an ENA emission source as viewed from the north ecliptic pole. The apparent ecliptic longitude of the source in the ram viewing $\lambda_r$ is larger than the heliocentric longitude $\lambda_0$. In the anti-ram viewing, the situation is opposite: $\lambda_a<\lambda_0$. The source is located at distance $d$ from the Sun. The arrows at the \emph{IBEX} symbols mark the directions of \emph{IBEX} motion around the Sun and of its spinning. The direction of the increase of the ecliptic longitude is the same as the ecliptic longitude of the Earth's motion and is indicated by an arrow.}
 \label{fig:parallax}
\end{figure}

\emph{IBEX} is moving relative to the Sun and each portion of the sky is observed twice per year, once when the spacecraft is moving towards it (which we will refer to as the ram viewing), and half a year later from the opposite side of the Sun, when the spacecraft is moving away from it (the anti-ram viewing). Figure~\ref{fig:parallax} presents a schematic view of the parallax shift of the ENA source according to the this observational strategy. The distance from the Sun $d$ to the ENA emission source can be calculated from the parallax angle $p$, obtained from the comparison between the source apparent position in the two viewings: 
\begin{equation}
 d=\frac{1\text{ AU}}{\tan p}. \label{eq:parallax}
\end{equation}

\subsection{The ribbon in the ram and anti-ram hemispheres}
\label{sec:ribbon}
\emph{IBEX} was launched into a highly elliptical orbit around the Earth, with a period of $\sim$7.5 days and apogee $\sim$50$R_\text{E}$ \citep{scherrer_etal:09a}. In mid-2011, the orbit was changed to a lunar-synchronized orbit with the orbital period of $\sim$9.1 day \citep{mccomas_etal:11a}. \emph{IBEX} is spinning at $\sim$4 rpm around its Z-axis, which approximately follows Sun, and the \emph{IBEX}-Hi and \emph{IBEX}-Lo instruments are mounted in the XY-plane. The Z-axis had been repositioned once per orbit near perigee during the period with the $\sim$7.5 days orbits, and has been repositioned twice per orbit, near the perigee and apogee, with the $\sim$9.1 days orbit \citep{mccomas_etal:14b}. This observational strategy allows probing the same swath of the sky twice per year from the opposite sides of the Sun (see Figure~\ref{fig:parallax}). This configuration provides the maximum possible parallax.

\emph{IBEX}-Hi is a single pixel ENA imager whose boresight is perpendicular to the \emph{IBEX} spin axis. Over a spacecraft spin, ENA measurements are acquired over a great circle perpendicular to the spin axis. ENAs are measured as a function of spin angle, where 0$^\circ$ corresponds to the point at which the great circle is closest to the north ecliptic pole (NEP), 90$^\circ$ corresponds to the spacecraft ram direction (i.e., the direction of motion of the spacecraft around the Sun), 180$^\circ$ is the closest point to the south ecliptic pole, and 270$^\circ$ is the anti-ram direction. We subsequently refer to the spin angle as the NEP angle as a reminder of its 0$^\circ$ reference toward the north ecliptic pole. Because of the motion of the Earth around the Sun, we must correct the measured ENA energy, flux, and angle-of-arrival based on the NEP angle relative to the Earth's velocity vector. To perform the velocity transformation to correct for this effect, we separately analyzed observed ENAs for NEP angles in the range $(0^\circ,\,180^\circ)$, to which we refer as the ram hemisphere, and in the range $(180^\circ,\,360^\circ)$ in the anti-ram hemisphere.

\begin{figure}[ht!]
\epsscale{1.}
 \plotone{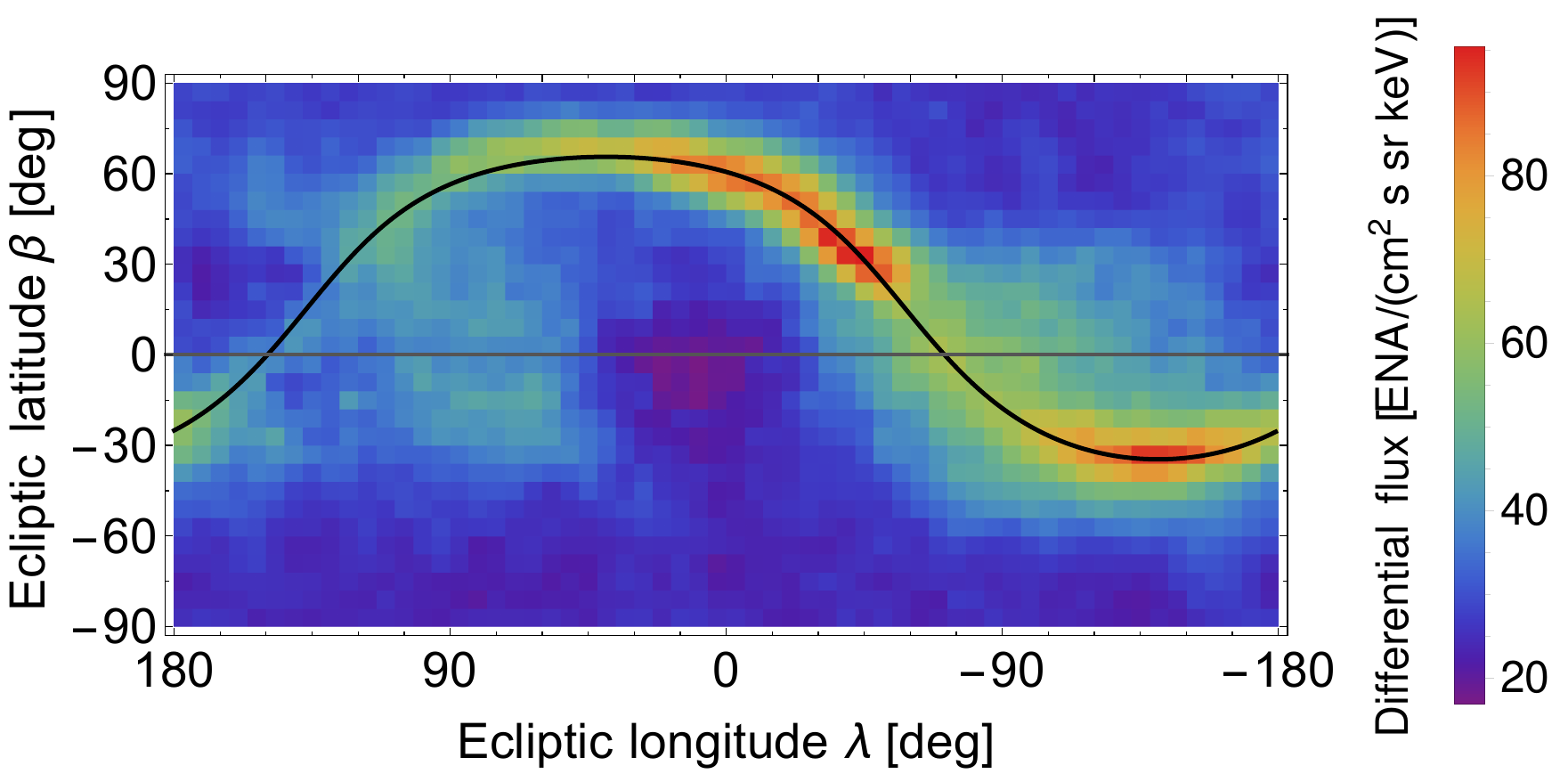}
 \caption{Observed all-sky ENA flux map for energy $\sim$1.7~keV based on the first 5-years of \emph{IBEX} data \citep{mccomas_etal:14b} in the ecliptic coordinate system with a equirectangular projection. The black curve presents the ribbon circle from \citet{funsten_etal:13a}, which traverses a unique latitude for each ecliptic longitude.}
 \label{fig:ribbonfun}
\end{figure}

Figure~\ref{fig:ribbonfun} presents the ENA flux map at 1.7 keV central energy based on the first 5-years of \emph{IBEX} data \citep{mccomas_etal:14b} in the equirectangular projection. In addition, the circle found by \citet{funsten_etal:13a} as an approximation of the location of maximum ribbon flux (hereafter the ribbon circle) is shown as the solid black line. This circle intersects each meridian in one point. Thus, ecliptic latitude of this point is an unambiguous function of ecliptic longitude. We denote this function as $\beta_\text{F}(\lambda)$. We note that \emph{IBEX}-Hi traverses the ribbon twice per spacecraft spin, once each in the ram and anti-ram hemispheres. For each orbital arc, we searched for the signal from the ribbon and looked for the NEP angle for which the signal from the ribbon is the maximum. This NEP angle was subsequently transformed to the apparent position in the ecliptic coordinates.

The \emph{IBEX}-Hi sensor is able to measure hydrogen atoms, and is equipped with an electrostatic analyzer (ESA) to select specific ENA energy passbands for measurement. Our analysis uses five of the six energy passbands, which have central energies 0.7, 1.1, 1.7, 2.7, and 4.3 keV and further on are referred to as ESA steps 2 to 6 \citep{funsten_etal:09a}.
The energy resolution is $\Delta E/E\sim 0.65$ FWHM. The acceptance directions are restricted by a collimator, with a field of view (FOV) of $\sim$6.5$^\circ$ FWHM. The collimator transmission function has a hexagonal symmetry to maximize the ENA throughput, but in this analysis we assume that the collimator transmission function has a cylindrical symmetry, which is substantiated by the analysis of \citet{sokol_etal:15b} of the nearly-identical \emph{IBEX}-Lo collimator.

\subsection{ENA trajectories}
After the ENAs are created, their motions are no longer governed by the electromagnetic forces, but by gravity and solar radiation pressure. The neutral atom trajectories can be obtained from integration of the equation of motion with these forces \citep{bzowski:08a}:
\begin{equation}
 \vec{F}(\vec{r},t,v_r)=-\frac{GM_\Sun m(1-\mu(v_r,t))}{r^2}\frac{\vec{r}}{r}\, , \label{eq:force}
\end{equation}
where $\vec{r}$ is the position of the atom in the Sun frame, $m$ is the atom mass, $M_\Sun$ is the Sun mass, and $G$ is the gravitational constant. The ratio of the solar resonant radiation pressure force to solar gravity $\mu$ depends on the time $t$ and the radial component of the atom's velocity ($v_r$). The equation of motion given by Equation~\eqref{eq:force} requires numerical solution due to the time and radial velocity dependence of $\mu$. Namely, the ratio $\mu$ depends on the Lyman-$\alpha$ line profile, which varies with time. Here we use the model by \citet{tarnopolski_bzowski:08a} with $\mu$  dependent on the total solar Lyman-$\alpha$ flux $I_\text{tot}$. The solar hydrogen Lyman-$\alpha$ line profile was adopted after \citet{bzowski_etal:13a}.

\begin{figure}[ht!]
\epsscale{0.7}
\plotone{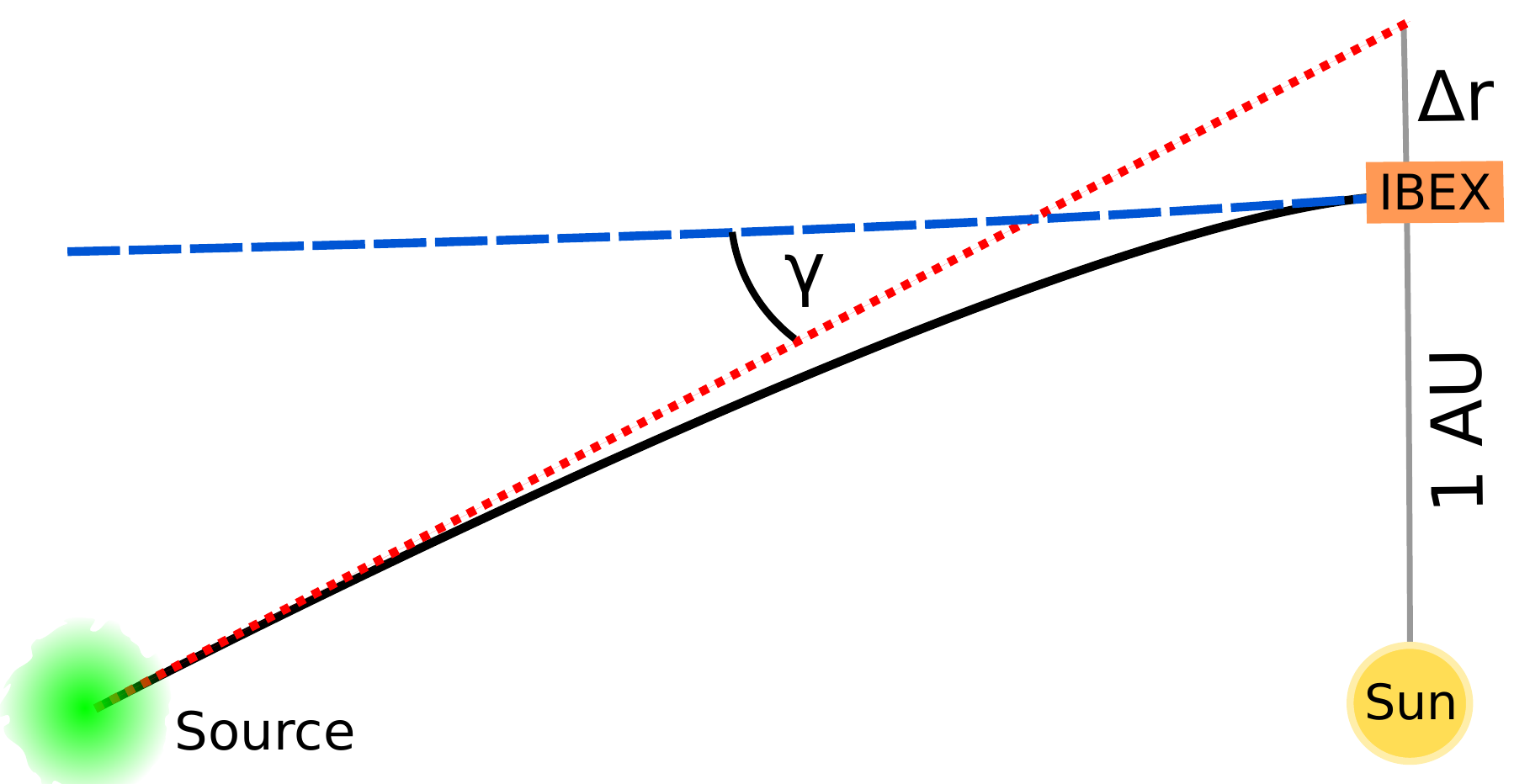}
 \caption{Illustration of ENA trajectories from the source to \emph{IBEX}. The actual trajectory (solid black line) is deflected by gravity and radiation pressure forces from the straight path trajectory (dotted red line). The apparent incoming ENA direction (dashed blue line) as measured at \emph{IBEX} is deflected by an angle $\gamma$. The gravity and radiation pressure change the distance at which the ENA crosses our parallax baseline by $\Delta r$.}
 \label{fig:trajectories}
\end{figure}

The gravity and radiation pressure are often neglected when the ENAs are analyzed, because the typical ENA velocities are an order of magnitude higher than the escape speeds. As a result, trajectories are often assumed to be straight lines. However, deflection by these forces could influence the parallax determination because the Sun, which lies at the midpoint of our parallax baseline, is also the object that governs the central-force behavior described by Equation~\eqref{eq:force}. Figure~\ref{fig:trajectories} illustrates the deflection of an ENA trajectory by these forces, described by an angle $\gamma$ and distance $\Delta r$. A quantitative assessment of this deflection for the geometries relevant for \emph{IBEX} observations is available in \citet{bzowski_tarnopolski:06a}. 

\begin{figure}[ht!]
\epsscale{0.7}
\plotone{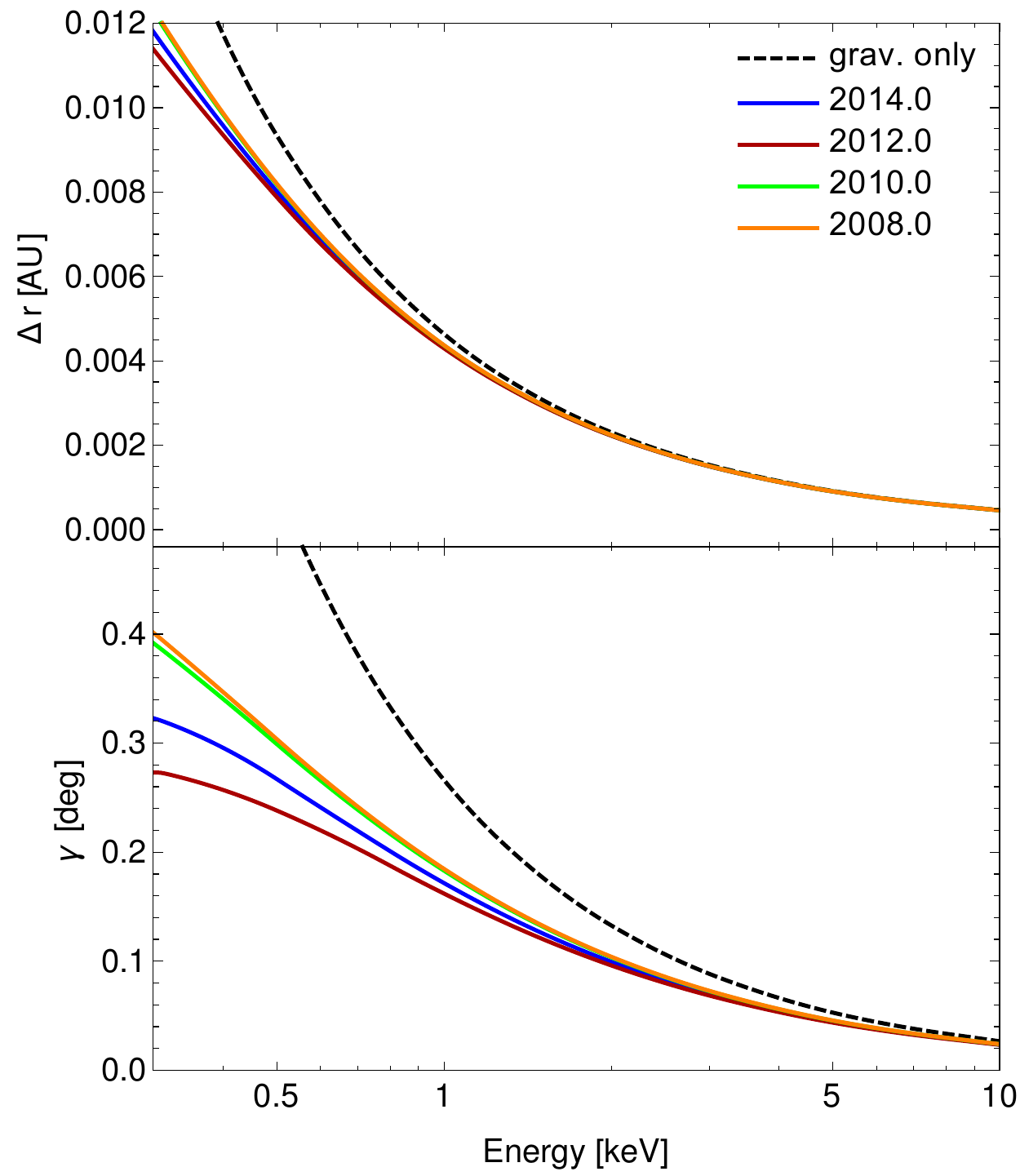}
 \caption{Modification of ENA trajectories due to solar gravity and radiation pressure, given as a distance $\Delta r$ along the parallax baseline (upper panel) and deflection angle $\gamma$ (lower panel) as measured at \emph{IBEX} as a function of ENA energy. The results for different times of arrival (solid lines) are compared to the deflection by gravity alone (dashed line), which of course does not depend on time.}
 \label{fig:deflection}
\end{figure}

Figure~\ref{fig:deflection} presents the magnitudes of the deflection depending on the ENA energy. The distance $\Delta r$, which potentially should be added to the parallax base, is small ($\lesssim 0.005$ AU) in comparison to the Sun-\emph{IBEX} distance of 1 AU. Consequently, we could neglect this effect when we calculate the distance to the source. However, the deflection angle $\gamma$ is of the same order as the potential parallax angle. Thus, the obtained apparent positions of the ridge need to be corrected for this effect in the analysis. Correction for the aberrated velocity (both direction and magnitude) of ENAs measured at \emph{IBEX} requires analysis of the ENA energy spectrum, which is described in the next section.

\section{DATA ANALYSIS}
\label{sec:data}
Often, \emph{IBEX} results are presented as sky maps of the ENA fluxes at various energies. In this analysis we used a different approach, where each orbital arc is analyzed separately to determine the apparent positions of the ribbon ridge, which were subsequently used to find the parallax shift. Determination of the ridge was done using the algorithm presented below. Results obtained in this analysis are based on the first 5 years of \emph{IBEX} data \citep{mccomas_etal:14b}, which cover the orbit range 11--230b.

\subsection{Step 1: Data histogram}
\label{sec:histogram}
\emph{IBEX}-Hi collects information about each measured atom in the form of the so-called Direct Events. For each event, the time of detection, the coincidence type, and the NEP angle are stored. The data are collected over the whole High Amplitude Science Operation time, but some time intervals must be removed due to various backgrounds \citep{mccomas_etal:12b, mccomas_etal:14b}. Selection of good observational conditions, the so-called ``good-times'' intervals, is a procedure strictly followed in the preparation of all \emph{IBEX} data products. In this analysis, we used the standard ``Good'' periods \citep{mccomas_etal:14b} with the additional restriction to the time intervals that do not require de-spinning.

In this analysis, each orbital arc and ESA step was analyzed separately. The events measured during the ``good-time'' intervals were histogrammed into 1$^\circ$-wide bins in NEP angle. Subsequently, the number of counts $n$ in each bin was divided by the time duration of the observation of this bin $\Delta t$. From the rate thus obtained, we subtracted the rate of the background $c_\text{bkg}$ due to the penetrating radiation \citep{mccomas_etal:14b} on this orbital arc and we divided the resulting rate by the survival probability $s$ \citep{mccomas_etal:14b} to get the corresponding corrected count rate $c$:
\begin{equation}
 c=\frac{1}{s}\left(\frac{n}{\Delta t}-c_\text{bkg}\right)\, . \label{eq:rate}
\end{equation}
The corrected count rates are proportional to the ENA flux measured in the \emph{IBEX} spacecraft frame. 

The histogram bins with widths of 1$^\circ$ are much narrower than the $\sim$6.5$^\circ$ FWHM of the \emph{IBEX}-Hi collimator response function. In fact, this wide \emph{IBEX}-Hi field of view causes some broadening of the observed features, but because of the scanning, the positions of the observed features, as the ridge apparent position, are not biased.

\subsection{Step 2: Compton-Getting correction for the flux}
\label{sec:cgflux}
\emph{IBEX} is moving with respect to the Sun, so the observed velocities of ENAs in the \emph{IBEX} frame are different than in the Sun frame, and the measured fluxes are modified due to the Compton-Getting effect. This effect is due to the invariance of the phase-space density of ENAs and has two important consequences. The first one is a difference between the apparent incoming direction of ENAs and the velocity direction in the Sun frame. The other one is a change in the absolute flux in the \emph{IBEX} frame. The reduced data from the \emph{IBEX} mission are routinely corrected for this effect using an efficient procedure \citep{mccomas_etal:10c}. In this analysis we follow this procedure, which we had to apply to the data selected for our analysis, which are histogrammed into 1$^\circ$ bins and have de-spinned intervals excluded. Below we provide a detailed description of the Compton-Getting corrections, because they are essential for an accurate determination of the parallax.

The ENA velocity in the Sun frame ($\vec{v}_\text{S}$, hereafter the subscript S denotes the Sun frame) is related to the velocity in the \emph{IBEX} frame ($\vec{v}_\text{I}$, hereafter the subscript I denotes the \emph{IBEX} frame) by the following formula:
\begin{equation}
 \vec{v}_\text{S}=\vec{v}_\text{I}+\vec{u}, \label{eq:velSI}
\end{equation}
where $\vec{u}$ denotes the spacecraft velocity in the Sun frame.

From the invariance of the phase-space density, the ENA flux in the Sun frame $j_\text{S}(\vec{\Omega}_\text{S},E_\text{S})$ is related to the flux in the \emph{IBEX} frame $j_\text{I}(\vec{\Omega}_\text{I},E_\text{I})$ by the following formula:
\begin{equation}
 j_\text{S}(\vec{\Omega}_\text{S},E_\text{S})=\frac{E_\text{S}}{E_\text{I}}j_\text{I}(\vec{\Omega}_\text{I},E_\text{I}), \label{eq:fluxframe}
\end{equation}
where $\vec{\Omega}$ represents the direction for which the flux is measured, and $E$ the energies of the observed ENAs. The ENA velocities are non-relativistic, thus $E_\text{S}/E_\text{I}=v_\text{S}^2/v_\text{I}^2$. In this paper, we follow the typical convention in which the direction of observation $\vec{\Omega}$ represents the instrument boresight, i.e., the direction opposite to the ENA velocities. The boresight in the \emph{IBEX} frame is retrieved from the NEP angle $\theta$, and the Z-axis pointing, which is fixed for each orbital  arc: $\vec{\Omega}_I=\vec{\Omega}_I(\theta,\lambda_o,\beta_o)$, where $(\lambda_o,\beta_o)$ are ecliptic coordinates of Z-axis at orbital arc $o$. The transformation to the Sun frame is calculated from Equation~\eqref{eq:velSI}.

The \emph{IBEX} velocity $\vec{u}$ can be decomposed into the velocity of the Earth around the Sun (speed $\sim$30~km~s$^{-1}$) and the \emph{IBEX} velocity relative to the Earth (speed $\sim$2~km~s$^{-1}$). \citet{sokol_etal:15b} found that the effect of the spacecraft velocity around the Earth is non-negligible for the observations of interstellar gas. In this paper, we also took the spacecraft motion around the Earth into account, because of the required high accuracy for the parallax determination, even though the velocities of the observed ENAs are much higher than those of the interstellar atoms. Here we analyzed the mean count rates over ``good-time'' intervals in each orbital arc. Thus, for the Compton-Getting correction we used the mean spacecraft velocity over these intervals on each orbital arc:
\begin{equation}
 \vec{u}_o=\langle \vec{u}(t) \rangle_{o,\text{GT}},
\end{equation}
where $\vec{u}(t)$ is the instantaneous velocity relative to the Sun and $\langle \dots \rangle_{o,\text{GT}}$ represents averaging over the ``good-time'' intervals. If the ``good-times'' intervals cover the whole \emph{IBEX} orbit then this average value is close to the Earth's velocity, but it is not always the case. Furthermore, after the lunar resonance orbit was established, the measurements are collected on an orbital arc (half-orbit) basis, and the average \emph{IBEX} velocity over each arc is slightly different from Earth's velocity. 

The energy response in each ESA step is defined in the \emph{IBEX} frame. Thus, the observed ENAs in the Sun frame have different energies depending on the NEP angle. We estimated the fluxes in the Sun frame for the fixed energies assuming that over a short energy range the fluxes could be approximated by a power law: $j\propto E^{-\gamma}$, where $\gamma$ is the spectral index. For convenience, we use the central energies $E_\text{I}$ for given ESA steps as the basis for normalization:
\begin{align}\nonumber 
 j_\text{S}(\vec{\Omega}_\text{S},E_\text{I})&=\left(\frac{E_\text{I}}{E_\text{S}}\right)^{-\gamma}j_\text{S}(\vec{\Omega}_\text{S},E_\text{S})\\&=\left(\frac{E_\text{I}}{E_\text{S}}\right)^{-\gamma-1}j_\text{I}(\vec{\Omega}_\text{I},E_\text{I}),
 \label{eq:fluxSunE}
\end{align}
where for the second equality we use Equation~\eqref{eq:fluxframe}. 

The observed count rates are proportional to the flux in the \emph{IBEX} frame: $c\propto j_\text{I}(\vec{\Omega}_\text{I},E_\text{I})$. To obtain a quantity proportional to the flux in the Sun frame, we defined the Compton-Getting (CG) corrected count rate:
\begin{equation}
 c_\text{CG}=\left(\frac{E_{\text{I}}}{E_{\text{S}}}\right)^{-\gamma-1}c, \label{eq:cgrate}
\end{equation}
where $E_{\text{I}}$ is the central energy in the considered ESA step, and $E_{\text{S}}$ is the energy in the Sun frame, calculated separately for each NEP angle. This correction requires the knowledge of the spectral index $\gamma$. As the first estimation of this value, we used spectral indices derived by \citet{desai_etal:15a} for the generally distributed flux (GDF) between the ESA steps as a function of ecliptic latitude. We estimated the values at the central energies using Equation~\eqref{eq:cgrate} and the interpolation and extrapolation of the spectral indices. The ribbon flux has a different energy spectrum from the GDF, thus the spectral index was updated after step 4 (Section~\ref{sec:spectra}) and the CG-corrected count rates were recalculated. For each ESA step the corrected count rate $c_\text{CG}$ is directly proportional to the flux and depends on the geometrical constant of considered ESA step.

\begin{figure*}[ht!]
\epsscale{1.0}
\plotone{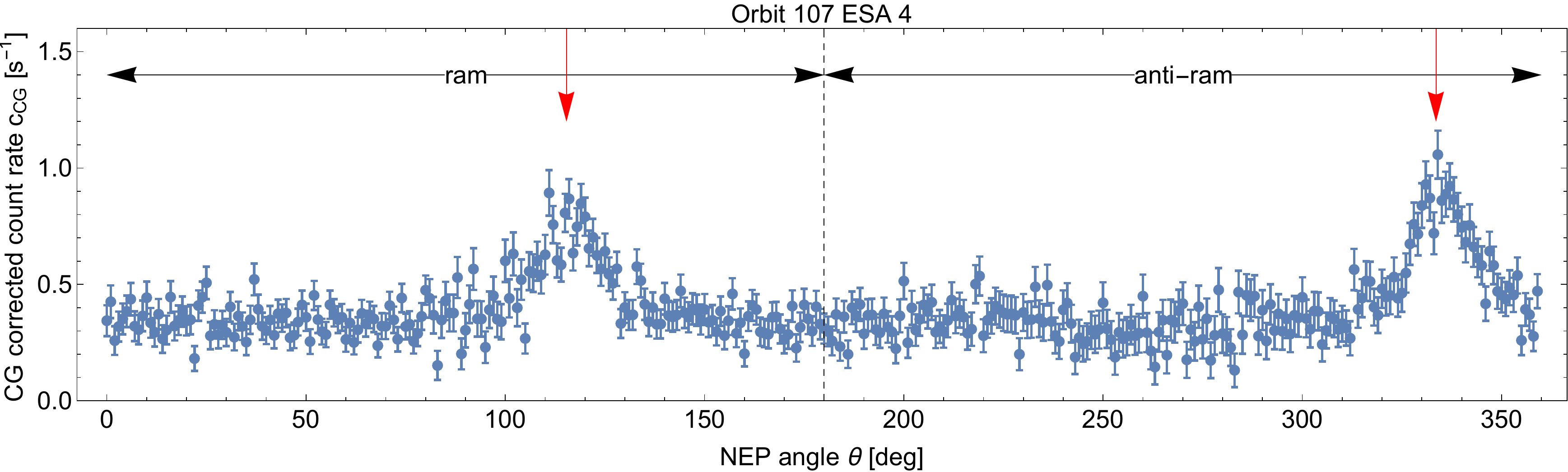}
 \caption{The CG-corrected count rate as a function of the NEP angle for orbit 104 in ESA step 4. The vertical line marks the boundary between the ram and anti-ram hemispheres, and the positions of intersections of the observed strip in the sky with the ribbon circle are indicted by the red arrows.}
 \label{fig:orbitdata}
\end{figure*}

In Figure~\ref{fig:orbitdata}, we present as an example the CG-corrected count rates given by Equation~\eqref{eq:cgrate} from orbit 107 in ESA step 4 as a function of the NEP angle. The mean rate outside the ribbon is $\sim$0.4 s$^{-1}$ and it increases to $\sim$1.0 s$^{-1}$ at the maximum of the ribbon. 

\subsection{Step 3: Fitting the ridge apparent position}
\label{sec:fitting}
For a given orbital arc, the ribbon circle is intersected twice by the instrument boresight, for two NEP angles, one in the ram hemisphere, and the other one in the anti-ram hemisphere (indicated by the red arrows in Figure~\ref{fig:orbitdata}). Since the ridge apparent positions should be close to these NEP angles, we used them for selection of the data segments for analysis. Namely, we selected two segments 70$^\circ$-wide, each centered at one of these NEP angles for each orbital arc and each ESA step. Some combinations of orbital arcs, ESA step, and hemisphere (ram or anti-ram) with the gaps in the NEP angles (possibly due to limitation of the ``good-time'' intervals to specified NEP angle intervals) were removed from the analysis. As a consequence, the number of fitted data points was $N=70$ for all of the analyzed segments. Additionally, we removed low statistics data with the accumulated time of observation for the segment less than 1000~s. 

The Compton-Getting corrected count rate $c_\text{CG}$ for the selected segments of NEP angles was subsequently fitted to follow the formula:
\begin{equation}
 f(\theta)=A+B(\theta-\theta_\text{r})+Ce^{-\frac{(\theta-\theta_\text{r})^2}{2\sigma^2}}\, , \label{eq:model}
\end{equation}
where $\theta$ is the NEP angle, and $A$, $B$, $C$, $\theta_\text{r}$, and $\sigma$ are the fit parameters. The fit parameter $\theta_\text{r}$ yields the apparent position of the ribbon ridge in NEP angle. In this formula, we assumed that the signal can be decomposed into the GDF and the ribbon flux portions, where GDF is linearly changing with the NEP angle, and the ribbon can be approximated by the Gaussian function.

The number of counts observed in a histogram bin generally follows the Poisson distribution if measured by an ideal detector \citep{olive_etal:14a}. This is true even if the count rates vary with time.  For \emph{IBEX}, the limited maximum throughput could potentially affect this, if a large number of counts hit the instrument in a short time \citep[see a throughput discussion in][]{swaczyna_etal:15a}. Such situations can be identified based on auxiliary data and are removed from the ``good time'' intervals. Thus the assumption that the observed numbers of counts in 1$^\circ$ bins during these intervals follow the Poisson distribution is well justified.

Since the observed numbers of counts in individual bins on some orbital arcs are relatively small (of the order of $\sim$10), we could not assume that the uncertainty is symmetrical and normally distributed. Consequently, using the least-squares method is not justified in this case. The problem of model fitting to low-count observations was discussed in the context of astrophysical results and high-energy physics data by, e.g., \citet{baker_cousins:84a}, \citet{mighell:99a} and \citet{hauschild_jentschel:01a}. As the most reliable solution we decided to adopt the minimum of the normalized Maximum Likelihood Estimator (MLE) in the form \citep{baker_cousins:84a}:
\begin{equation}
 \text{MLE}(\vec{\pi})=2\sum_{i=1}^{N} \left[ \mu_i(\vec{\pi})-n_i+n_i\log\frac{n_i}{\mu_i(\vec{\pi})}\right], \label{eq:mle}
\end{equation}
where $\mu_i(\vec{\pi})$ is the expected number of counts from the model with parameters $\vec{\pi}$, and $n_i$ is the observed number of counts. The sum is taken over all used data points. The last term is counted as 0 if $n_i=0$. 

The expected number of counts for the model given by Equation~\eqref{eq:model} could be obtained using Equations~\eqref{eq:cgrate} and \eqref{eq:rate}:
\begin{equation}
 \mu=\Delta t\left[c_{\text{bkg}}+s\left(\frac{E_{\text{I}}}{E_{\text{S}}}\right)^{\gamma+1}\left[ A+B(\theta-\theta_\text{r})+Ce^{-\frac{(\theta-\theta_\text{r})^2}{2\sigma^2}} \right]\right]\, .
\end{equation}
The model has 5 fit parameters for each orbital arc, ESA step, and hemisphere (ram or anti-ram). The best fit parameters ($\vec{\pi}_0$) were obtained by numerical minimization of the estimator given in Equation~\eqref{eq:mle}. The related covariance matrix could be obtained from the matrix of second derivatives of the estimator at the minimum:
\begin{equation}
 \mat{\Sigma}=\left(\frac{1}{2}\left.\frac{\partial^2 \text{MLE}(\vec{\pi})}{\partial\vec{\pi}^2}\right|_{\vec{\pi}=\vec{\pi}_0}\right)^{-1}\, .
 \label{eq:corrmat}
\end{equation}
The uncertainties of the respective parameters could be obtained as square root of the related diagonal element of covariance matrix $\mat{\Sigma}$.

\begin{figure*}[ht!]
\epsscale{1.0}
\plottwo{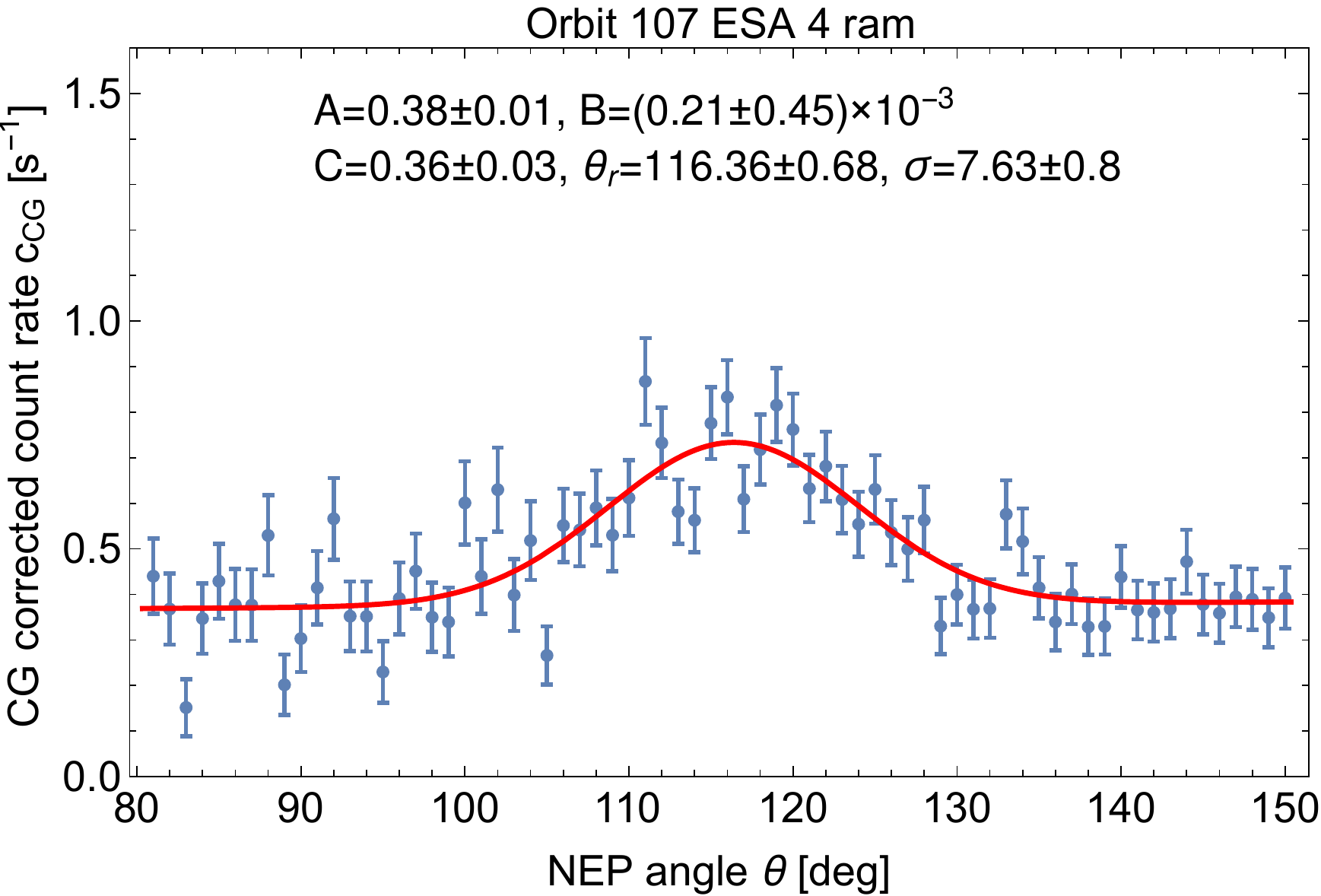}{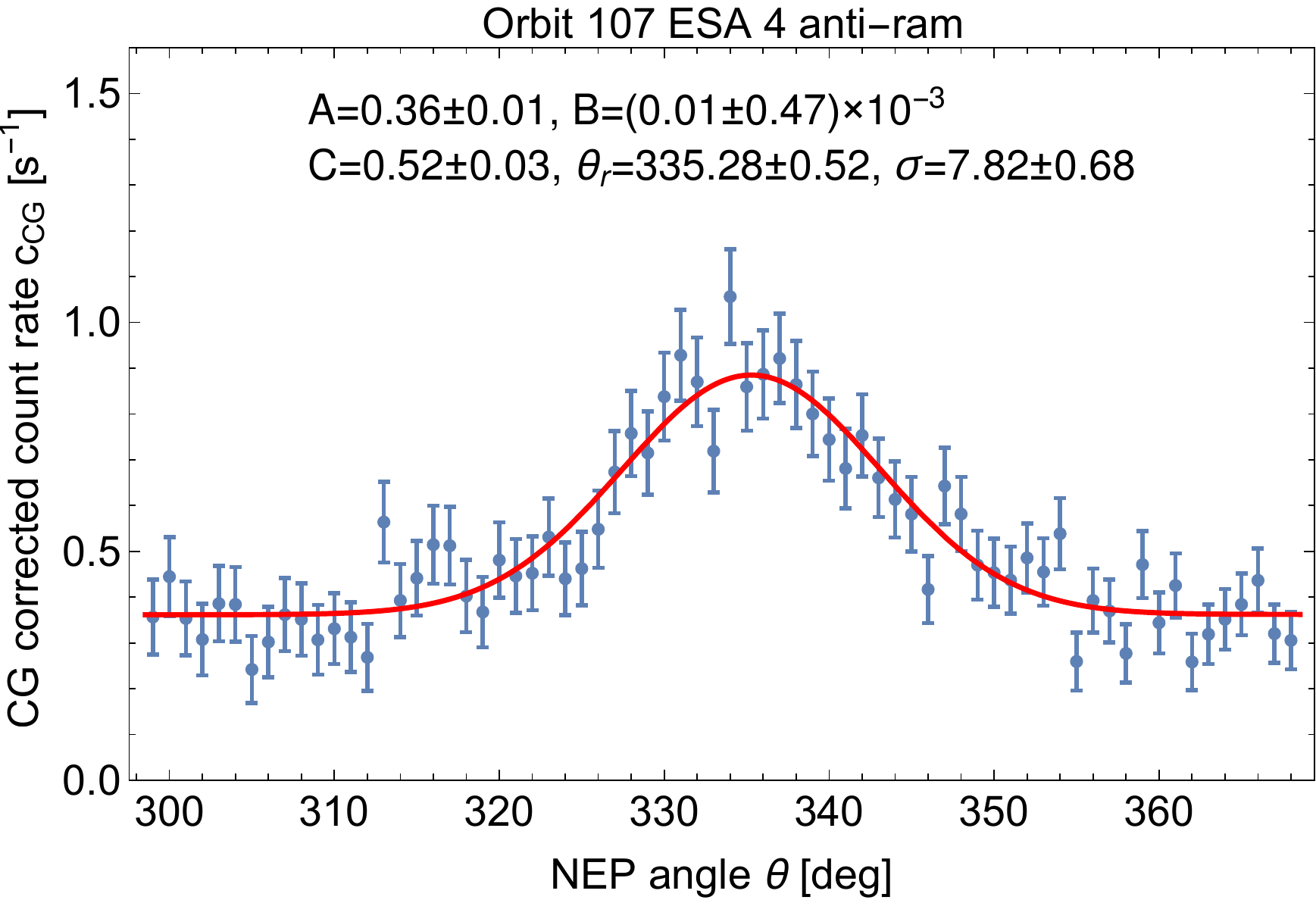}
 \caption{CG-corrected counts rates for segments selected for fitting in the ram (left panel) and anti-ram (right panel) hemispheres. The red solid lines present the obtained fits. The fitted parameters are presented in the upper part of the figures.}
 \label{fig:orbitfits}
\end{figure*}

The model given by Equation~\eqref{eq:model} is non-linear in the fit parameters. Namely, the fitted position of the ridge $\theta_\text{r}$ and the width of the ribbon $\sigma$ could be estimated incorrectly if the height of the ribbon $C$ does not significantly differ from zero. We select only the  fits where $C>2.5\delta C$, where $\delta C$ is the uncertainty of the parameter $C$. It is statistically expected that this condition will be accidentally fulfilled by about 1 case in 100 for segments that do not contain the ribbon signal. Additionally, we require that the width $\sigma$ of the flux peak lie in the range $3^\circ<\sigma<25^\circ$. We adopted these criteria to avoid non-physical peak identification in the regions where the ribbon flux is poorly visible.
Figure~\ref{fig:orbitfits} presents the obtained fits for the data from orbit 107, ESA step 4. 

\begin{figure*}[ht!]
\epsscale{1.0}
\plottwo{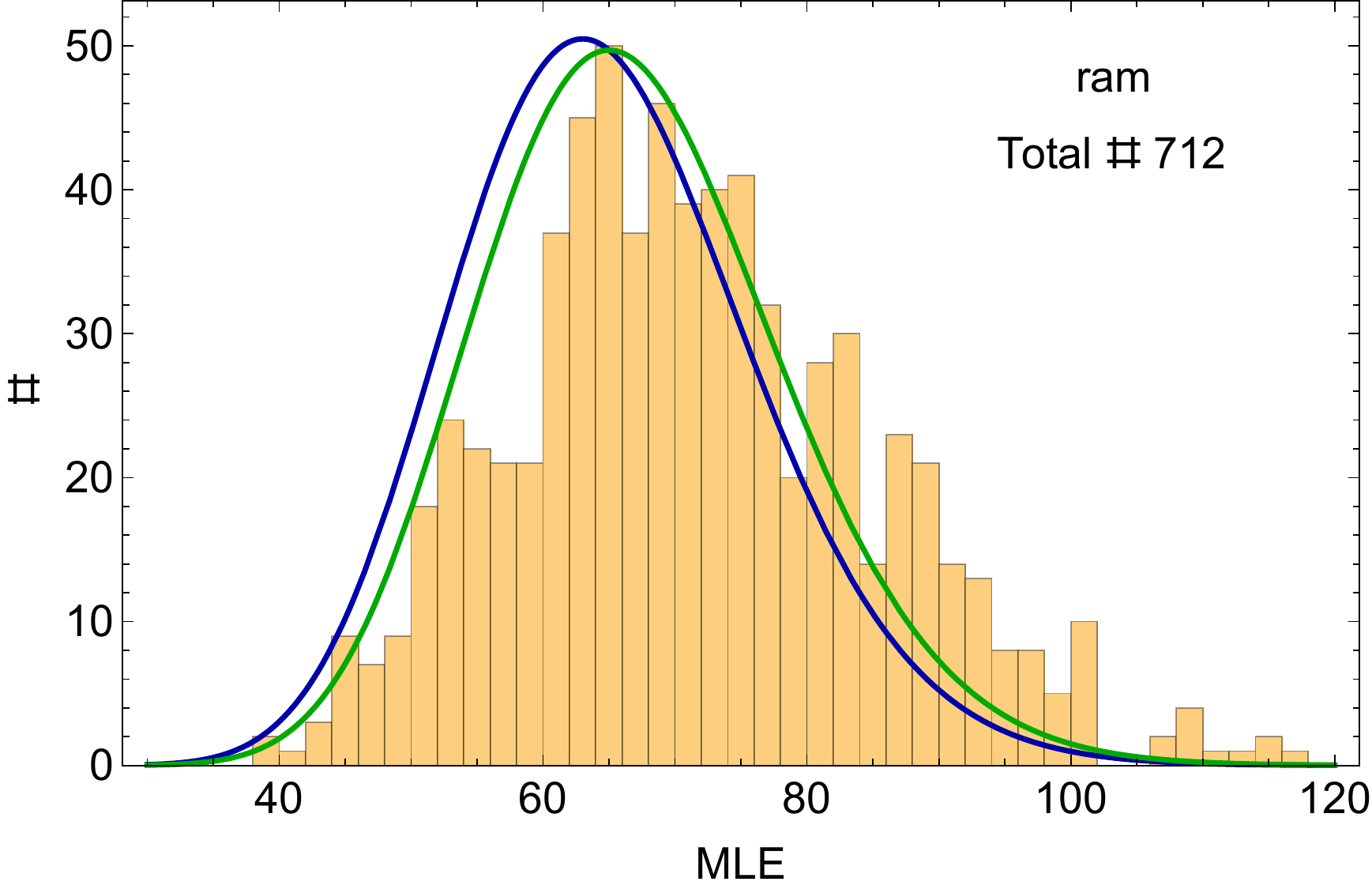}{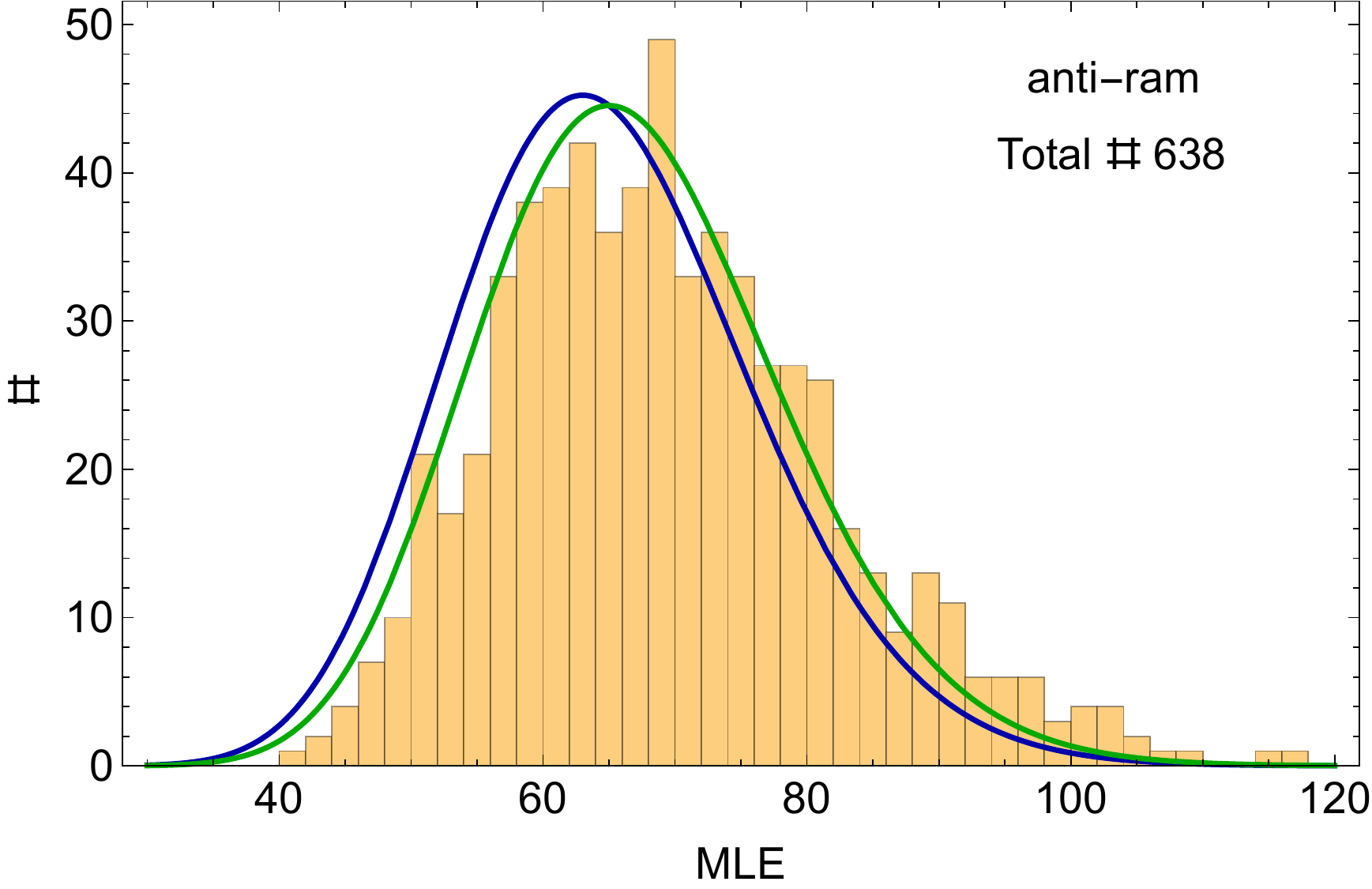}
 \caption{Histograms of the MLE values from Equation~\eqref{eq:mle} at minimum for fits to the data in the ram (left panel) and anti-ram (right panel) hemispheres. Only the fits with a significant signal from the ribbon are presented (cf. text). The $\chi^2$-distributions with 65 and 67 degrees of freedom (blue and green lines respectively) are presented for comparison (see discussion in the text).}
 \label{fig:histograms}
\end{figure*}

The minimum values of the estimator $\text{MLE}(\vec{\pi})$ should be distributed as a $\chi^2$-distribution. In Figure~\ref{fig:histograms} we present histograms of the minimum MLE values for fits conforming with the conditions discussed above. For a non-linear fit, determination of the number of degrees of freedom is not simple. The problem of non-linear fits to the data is discussed by \citet{andrae_etal:10a}. In the presented models, 3 parameters ($A$, $B$, and $C$) are linear and 2 parameters ($\sigma$ and $\theta_\text{r}$) are not linear. This follows from inspection of the limit $C=0$, for which the corrected flux given in Equation~\eqref{eq:model} does not depend on the values of the parameters $\sigma$ and $\theta_\text{r}$. Consequently, the number of degrees of freedom must be between 65 ($=70-5$) and 67 ($=70-3$). We present both these theoretical distributions in Figure~\ref{fig:histograms}. The agreement between the MLE histograms and the $\chi^2$-distributions are reasonable and the differences could arise from the model imperfection. Still, the method used is sufficient to determine the ridge location needed in this analysis.

\citet{funsten_etal:13a} found that the ribbon profile in the ribbon centered frame is not symmetric, and the skewness is $\sim$1. In this analysis, we neglected the asymmetry of the ribbon, because the potential offset of the ridge positions in the ram and anti-ram hemisphere due to fitting of a symmetrical function should occur in the same direction on the sky, so it should not be a source of systematic bias. Additionally we neglected the uncertainty of the background count rate $c_\text{bkg}$ in the fitting, because it is much smaller than the statistical scatter. Both these issues could be potentially responsible for the difference of the obtained MLE histograms (Figure~\ref{fig:histograms}) from the expected theoretical distributions.

\subsection{Step 4: Determination of the energy spectra of the ribbon}
\label{sec:spectra}
In step 2 of the analysis (Section~\ref{sec:cgflux}), we corrected the observed count rates by normalizing them to the central energy of the \emph{IBEX}-Hi ESA steps. For this, we needed the spectral indices for each bin, and we used only the GDF indices derived by \citet{desai_etal:15a}. In this section, we modify the spectral indices by the contribution from the ribbon flux. 

The flux varies along the ribbon differently in each ESA step \citep{schwadron_etal:14b, funsten_etal:15a}. Consequently, for assessment of the ribbon flux we divided the ribbon into 15 equally long ($\Delta \lambda=24^\circ$) sectors depending on the ecliptic longitude of the ribbon $\lambda$. In each sector, we assessed the ribbon spectrum based on the fitted flux maximum $c$ of the ribbon and the energy geometric factors \citep{funsten_etal:09a}. The spectra obtained were subsequently used to assess the spectral indices calculated between adjacent energy steps. 

\begin{figure}[ht!]
\epsscale{0.55}
\plotone{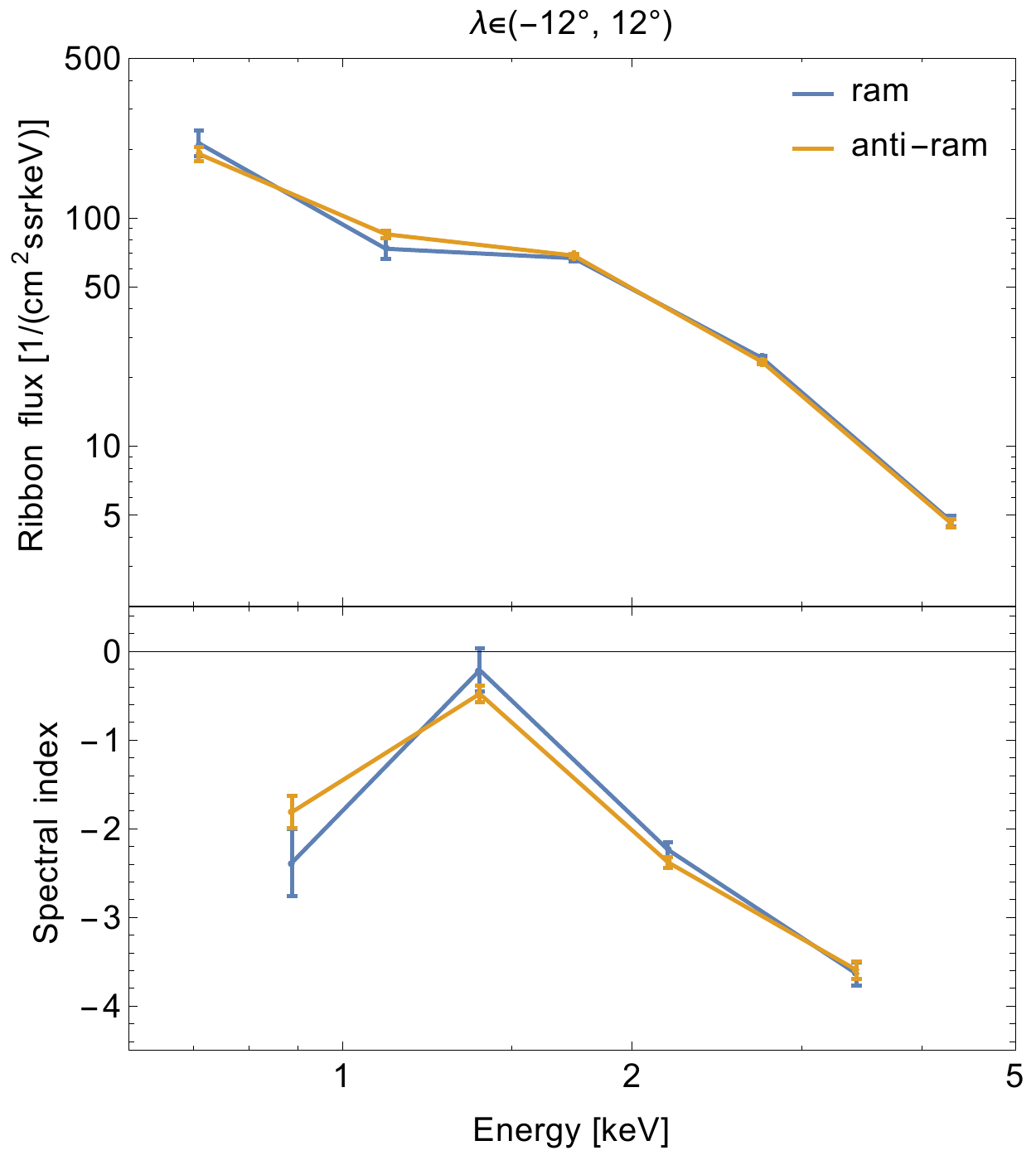}
 \caption{ENA spectra (upper panel) and the spectral indices (bottom panel) of the ribbon for the ecliptic longitude $\lambda\in(-12^\circ,12^\circ)$. Presented are spectra obtained from the ram and anti-ram viewing.}
 \label{fig:ribbonspectra}
\end{figure}

As an example, in Figure~\ref{fig:ribbonspectra} we present the  spectra and spectral indices obtained for the sector $\lambda\in(-12^\circ,12^\circ)$. Notice that the spectra obtained from the ram and anti-ram measurements closely agree.  

Using these ribbon spectral indices in place of the spectral indices used in Section~\ref{sec:cgflux} as given by the weighted mean with weights equal to the appropriate contributions to the total signal from the GDF and the ribbon flux (see Equation~\eqref{eq:model}) in each NEP angle separately:
\begin{equation}
 \gamma=\frac{\left[A+B(\theta-\theta_\text{r})\right]\gamma_{\text{GDF}}+C\exp\left(-\frac{(\theta-\theta_\text{r})^2}{2\sigma^2}\right)\gamma_{\text{R}}(\lambda)}{A+B(\theta-\theta_\text{r})+C\exp\left(-\frac{(\theta-\theta_\text{r})^2}{2\sigma^2}\right)}\, , 
\end{equation}
where $\gamma_{\text{GDF}}$ is the GDF index, and $\gamma_\text{R}(\lambda)$ is the ribbon index. We obtained the spectral indices at the central energies from the linear interpolation or extrapolation for the lowest and highest energy steps. 

\subsection{Step 5: Corrections of the ridge direction}
\label{sec:corrdir}
Based on the NEP angles of maximum ribbon flux ($\theta_\text{r}$) obtained in Section~\ref{sec:fitting}, we calculated the incoming direction of ENAs on the sky and corrected this direction for the Compton-Getting effect as given by Equation~\eqref{eq:velSI}. The ENA speed was calculated from the mean energy of the ribbon ENAs observed in each energy step. For this, we utilized the response function of the \emph{IBEX}-Hi sensor presented by \citet{funsten_etal:09a}, which we convolved with the observed spectrum of the ribbon. The resulting mean energy derived from this convolution can substantially deviate from the mean energy of the ESA step because of the high value of the spectral index, especially at the higher energies. 

After Compton-Getting correction, the velocities of the ENAs along the ridge of the ribbon are known in the Sun frame. As discussed in Section~\ref{sec:trajectories}, deflection due to gravity and radiation pressure are co-planar with the parallax measurement. As already noted, the change of the parallax base length is negligible ($\lesssim0.004$ AU), thus we could neglect this effect in the analysis. However, we corrected the apparent position of the ridge by the deflection angle calculated from the numerical solution of the equation of motion with the time of arrival equal to the mean observation time on each orbital arc. 

With the procedure described in Sections~\ref{sec:histogram}--\ref{sec:corrdir}, we obtained the NEP angles ($\theta_\text{r}$) of the ribbon ridge and their uncertainties (via Equation~\eqref{eq:corrmat}) for individual orbits and ESA steps, separately for the ram and anti-ram hemispheres. Moreover, the spin axis is located almost exactly in the ecliptic plane and known accurately \citep{swaczyna_etal:15a}. We transformed the obtained NEP angle to ecliptic coordinates and their uncertainties have dominant components along ecliptic latitude, whereas the uncertainties in ecliptic longitude are negligible. In the following analysis, we assume that the uncertainties of the apparent ridge positions are only in the ecliptic latitude.

\section{THE RIBBON PARALLAX}
Using the procedure described in Section~\ref{sec:data}, we obtained the apparent position of the ridge for each ESA step and hemisphere (ram or anti-ram) from the selected orbital arcs. We expect that a systematic shift of these apparent positions between the ram and anti-ram hemispheres is a signature of a parallax effect. 

The apparent position of any object on the sky in the ecliptic coordinates is shifted due to parallax from its heliocentric position $(\lambda,\beta)$ by \citep{smart:77}:
\begin{align}
 \Delta\lambda \cos\beta &= p \sin (\lambda_\Sun-\lambda),\\
 \Delta\beta &= -p \sin\beta\cos(\lambda_\Sun-\lambda), 
\end{align}
in longitude and latitude, respectively. The parallax angle $p$ is given by Equation~\eqref{eq:parallax} and $\lambda_\Sun$ is the Sun longitude. In the ram viewing, the longitude difference is  $\lambda_\Sun-\lambda\approx90^\circ$, whereas in the anti-ram viewing it is $\lambda_\Sun-\lambda\approx-90^\circ$ (see Figure~\ref{fig:parallax}). For these angles cosine gives 0, so the shift in latitude vanishes. The ribbon latitudes differ from $90^\circ$, thus the shift in the ecliptic coordinates could be expressed as follows:
\begin{equation}
 (\lambda,\beta)\rightarrow\left(\lambda\pm \frac{p}{\cos\beta},\beta\right)\, , \label{eq:parallaxshift}
\end{equation}
where the $+$ sign is for a parallax shift in the ram hemisphere and the $-$ sign is for the anti-ram hemisphere.

\subsection{Selection of the ribbon segment}
\label{sec:selsegment}
The location of and distance to the emission source of the ribbon ENAs relative to the Sun may have a spectral dependence, as possibly indicated by a slightly different ribbon peak locations at different energies derived using ram flux maps \citep{funsten_etal:13a}.
In this analysis we confined the data to the range where the energy dependence is not visible. Otherwise, the comparison between the apparent positions in the ram and anti-ram hemispheres would require correction for this effect, because the observed atom energies in the ram and anti-ram hemispheres are slightly different due to the Compton-Getting effect. In that case, the apparent positions are shifted by the energetic dependence in addition to any parallax. However, derivation of such corrections is beyond the scope of this analysis.

\begin{figure}[ht!]
\epsscale{0.47}
\plotone{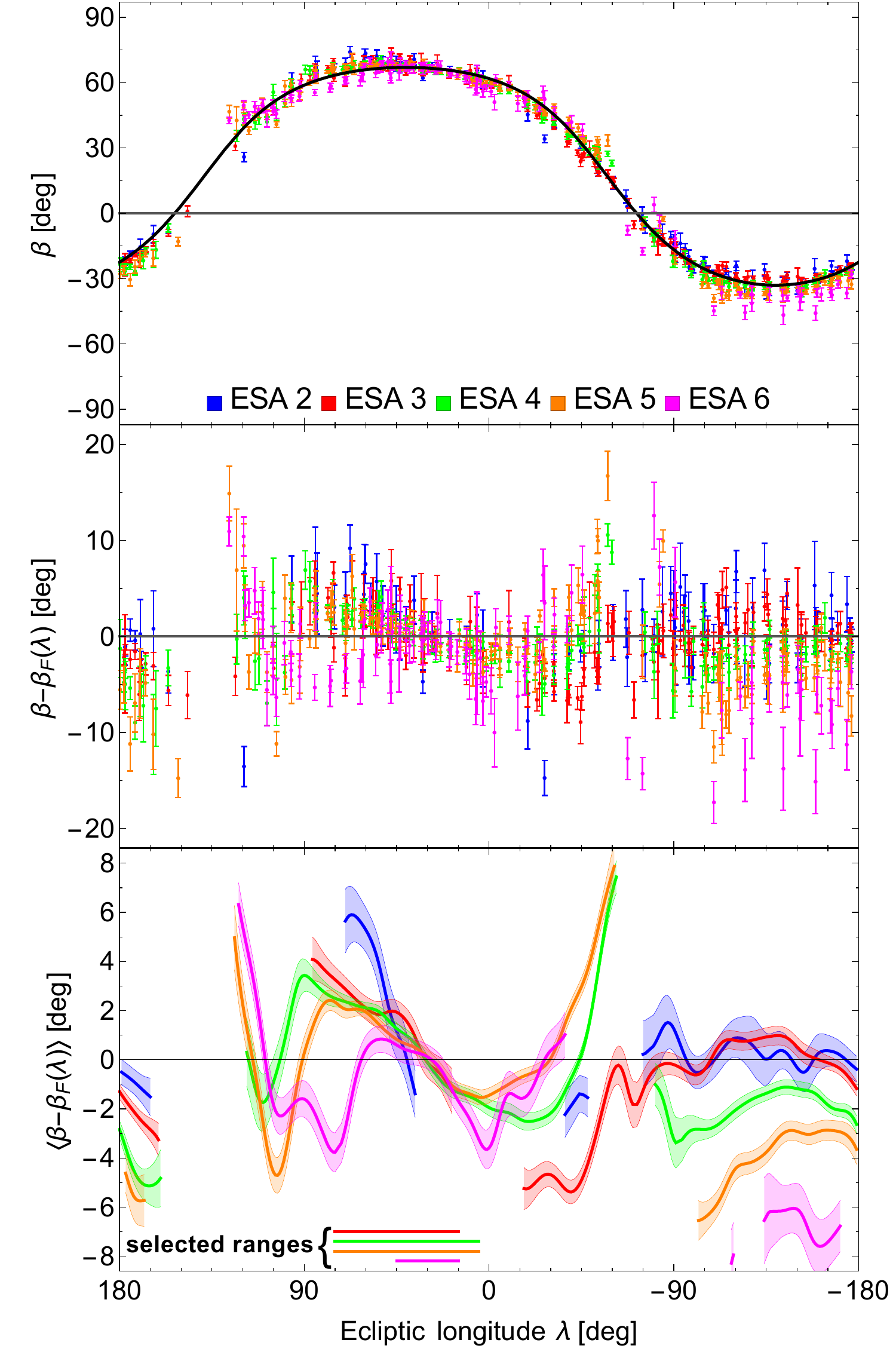}
 \caption{
 Comparison of the ridge apparent position among ESA steps 2--6 in the ram hemisphere. Upper panel: Apparent positions of the ridge from individual orbital arcs, corrected as described in Section~\ref{sec:data}, with the ribbon circle from \citet{funsten_etal:13a} superimposed. Middle panel: Arithmetic differences between the ridge ecliptic latitudes $\beta$ and the latitudes of the corresponding points on the ribbon circle $\beta_\text{F}(\lambda)$. Lower panel: The gaussian-smoothed differences from the middle panel, with the uncertainty bands added. The ranges selected for further analysis are presented in the bottom of the lower panel. Gaps represent the lack of a good fit for the ridge apparent position for corresponding energy and longitude. Notice the scale differences between the panels.
 }
 \label{fig:meanram}
\end{figure}

Figure~\ref{fig:meanram} compares the apparent positions of the ridge across all individual ESA steps. The absolute latitude are shown in the upper panel, and in the middle panel we replot them with the ecliptic latitude of the ribbon circle subtracted to reduce the large span of ecliptic latitudes and facilitate recognizing differences between the ribbon latitudes for different ESA steps. The differences gaussian-smoothed are showed with their uncertainty bands in the lower panel of Figure~\ref{fig:meanram}. One can easily observe that different energy steps have a common part for ecliptic longitudes in the range $\sim(0^\circ,90^\circ)$. 

From inspection of Figure~\ref{fig:meanram} one can find that the ridge apparent positions in ESA steps 4 and 5 agree with each other over the range of $\lambda\in(5^\circ,\,75^\circ)$. The apparent position of the ridge in ESA step 3 also conforms with the ESA steps 4 and 5, but the ridge was fitted only in a part of this range $\lambda\in(15^\circ,\,75^\circ)$. In ESA step 6, the apparent positions are in agreement only for a shorter range $\lambda\in(15^\circ,\,45^\circ)$. To avoid the potential bias, we restricted the analysis to these ranges. The ribbon is barely visible in ESA step 2 for this range, and the shape of the ridge apparent position is different, thus we do not use data from this step.

Additionally, we checked the consistency of the ram and anti-ram data in selected ranges for statistical outliers. Almost all data points are consistent with the moving average at a 3$\sigma$ level, except for 2 points in ESA step 4. 

\subsection{Parallax fitting}
\label{sec:parallax}
In the previous sections, we identified, quantified, and corrected for known systematic effects the observations of the location of the ribbon ENA emission source. The remaining systematic differences in the ridge locations between the ram and anti-ram maps are now attributed to the parallactic viewing of the emission source.

The heliocentric ridge position (i.e., without parallax shift) as projected onto the sky can be represented by a curve. Furthermore, we assume that ecliptic latitude of points on this curve is an unambiguous function of ecliptic longitude. The measured ridge apparent positions should fit to this curve after shifting them by the parallax angles in the appropriate directions for the ram and anti-ram hemisphere observations. We also expect that this function is smooth, so we are able to expand it locally into a polynomial.
We found that to obtain a statistically acceptable fit, we need at least a quadratic polynomial. Thus, the curve describing the ridge location is given by the formula:
\begin{equation}
 \beta(a_2,a_1,a_0;\lambda)=a_2\lambda^2+a_1\lambda+a_0\, , \label{eq:curve}
\end{equation}
where $a_2$, $a_1$, and $a_0$ are fitted parameters and $\lambda$ is ecliptic longitude. The least-square estimator $\chi^2$ for the parallax fit is:
\begin{align}
 \chi^2(p,a_2,a_1,a_0)&=\sum_{i=1}^{n_\text{r}}\left(\frac{\beta_i-\beta(a_2,a_1,a_0;\lambda_i-p\cos\beta_i)}{\delta\beta_i}\right)^2\nonumber\\
 &+\sum_{j=1}^{n_\text{a}}\left(\frac{\beta_j-\beta(a_2,a_1,a_0;\lambda_i+p\cos\beta_i)}{\delta\beta_j}\right)^2,
 \label{eq:chi2}
\end{align}
where we compare the ecliptic latitudes of the ridge obtained in Section~\ref{sec:data} ($\beta_{i,j}\pm\delta\beta_{i,j}$) with the location obtained from the curve given by Equation~\eqref{eq:curve} at the ecliptic longitude ($\lambda_{i,j}$) corrected to compensate for the parallax shift using the formula from Equation~\eqref{eq:parallaxshift}. The first sum in Equation~\eqref{eq:chi2} accounts for the apparent positions obtained in the ram hemisphere, and the second sum for the apparent positions from the anti-ram hemisphere. Consequently, the model has 4 fitted parameters: the parallax angle $p$ and the three parameters of the polynomial. However, we were interested mostly in the parallax angle $p$, and thus we performed minimization with respect to the curve parameters: $\chi^2(p)=\min_{a_0,a_1,a_2}\chi^2(p,a_2,a_1,a_0)$.

In Table~\ref{tab:fitted} we present the results of the fitting to the data from different ESA steps and also to their combinations. We provide the obtained minimum  of $\chi^2$ for each case, and the related number of degrees of freedom $\nu$. We checked the goodness of the fit and list the statistical $p$-value. We found that this value is smaller than 0.05, which implies a bad fit, only for those cases which included the two outliers from ESA step 4, mentioned in Section~\ref{sec:selsegment}. However, the parallax angles in the corresponding cases do not change significantly. This led us to the conclusion that inclusion of the outliers does not bias the results. Nevertheless, as the final result we quote the cases obtained without the outliers.

\begin{deluxetable}{lcrrrrrrrc}
 \tablewidth{0pt}
 \tablecolumns{6}
 \tablecaption{Fitting results \label{tab:fitted}}
 \tablehead{
    \colhead{}		&
    \colhead{}			&
    \multicolumn{4}{c}{Fitted parameters}	&
    \multicolumn{3}{c}{Fit quality}		&
    \colhead{} \\
        \cmidrule(lr{.75em}){3-6} \cmidrule(lr{.75em}){7-9} 
    \colhead{ESA}		&
    \colhead{distance (AU)}			&
    \colhead{parallax $p$ ($^\circ$)}		&
    \colhead{$a_0$} &
    \colhead{$a_1$} &
    \colhead{$10^3a_2$} &
    \colhead{$\nu$}				&
    \colhead{$\chi^2_\text{min}$}	&
    \colhead{$p$-value}	&
    \colhead{Fig.~\ref{fig:deltachi2}} 
 }
 \startdata
 3 			& $182^{+\infty}_{-112}$   	& \text{0.314$\pm $0.495} & 55.66 & -0.528 & 5.38  & 48  & 55.0  & 0.23	& \checkmark 	\\
 4 			& $262^{+\infty}_{-132}$  	& \text{0.218$\pm $0.221} & 59.65 & -0.336 & 3.29  & 89  & 98.1  & 0.24	& \checkmark 	\\
 4\tablenotemark{a} 	& $279^{+\infty}_{-144}$ 	& \text{0.205$\pm $0.218} & 59.62 & -0.339 & 3.32  & 91  & 120.3 & 0.02	& 	 	\\
 5 			& $\phantom{0}93^{+62\phantom{0}}_{-27}$ 	 	& \text{0.615$\pm $0.245} & 60.21 & -0.323 & 3.27  & 91  & 88.5  & 0.56	& \checkmark	\\
 6 			& $\phantom{0}76^{+293}_{-34}$ 	 	& \text{0.755$\pm $0.599} & 53.98 & -0.824 & 12.65 & 31  & 31.5  & 0.44	& \checkmark	\\
 3-6 			& $140^{+84\phantom{0}}_{-38}$ 		& \text{0.409$\pm $0.153} & 59.85 & -0.334 & 3.33  & 271 & 293.3 & 0.17	& \checkmark	\\
 3-6\tablenotemark{a} 	& $144^{+89\phantom{0}}_{-40}$		& \text{0.398$\pm $0.152} & 59.83 & -0.335 & 3.34  & 273 & 315.9 & 0.04	& 	 	\\
 3-5 			& $148^{+102}_{-43}$ 		& \text{0.386$\pm $0.157} & 59.85 & -0.334 & 3.33  & 236 & 256.0 & 0.18	& 	 	\\
 4-6 			& $138^{+86\phantom{0}}_{-38}$ 		& \text{0.415$\pm $0.159} & 59.92 & -0.329 & 3.27  & 219 & 232.0 & 0.26	& 	 	\\
 4-5 			& $146^{+105}_{-43}$ 		& \text{0.391$\pm $0.163} & 59.92 & -0.329 & 3.26  & 184 & 194.7 & 0.28	& 	 	

 \enddata
 \tablecomments{The column 'Fig.~\ref{fig:deltachi2}' indicates the cases presented in Figure~\ref{fig:deltachi2}.  $+\infty$ as an upper uncertainty of the distance means that the parallax angle is consistent with 0 at the 1$\sigma$ level, i.e., the distance is consistent with $\infty$. }
 \tablenotetext{a}{fits with the outliers included (Section~\ref{sec:selsegment})}
\end{deluxetable}

\begin{figure}[ht!]
\epsscale{0.7}
\plotone{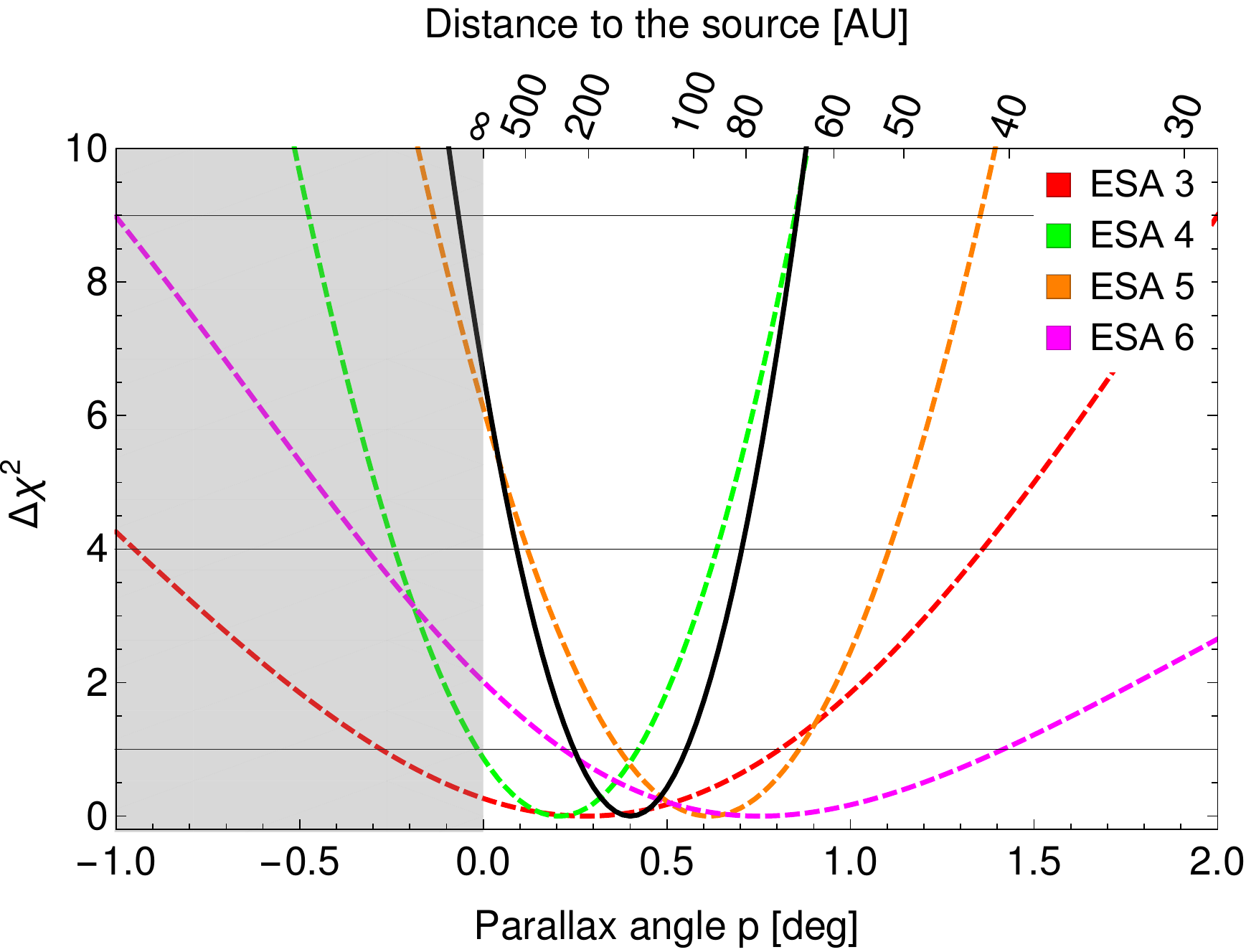}
 \caption{Difference of the estimator value from the minimum $\Delta\chi^2(p)=\chi^2(p)-\chi^2_\text{min}$ as a function of the parallax angle $p$. The acceptable regions at 1$\sigma$, 2$\sigma$, and 3$\sigma$ are bounded by $\Delta\chi^2=$ 1, 4, and 9, respectively. The black solid line presents the results from ESA steps 3-6. The dashed lines present the results of fitting to individual ESA steps. The shaded area presents the region of unphysical results ($p<0$). }
 \label{fig:deltachi2}
\end{figure}

Figure~\ref{fig:deltachi2} presents the acceptable ranges of the parallax angle $p$ based on the difference of the estimator value and the value at the minimum $\Delta\chi^2(p)=\chi^2(p)-\chi^2_\text{min}$ for the selected cases with 1$\sigma$ confidence intervals. The results from individual ESA steps are consistent with the result from all ESA steps analyzed together. Thus, the best fit parallax is $p=0.41^\circ\pm0.15^\circ$, which is equivalent to a distance of $140^{+84}_{-38}$~AU at the 1$\sigma$ confidence interval.

\section{DISCUSSION}
\label{sec:discussion}
Based on the data collected by \emph{IBEX}, we determined a parallax shift of $p=0.41^\circ\pm0.15^\circ$ for the ribbon. This offers a model-independent measurement of the distance of $d=140^{+84}_{-38}$~AU to the center of the ribbon source. We were able to obtain this result using a specially developed procedure for correction of the apparent position of the ribbon ridge. Based on most models of the ribbon, the ENA emission region is radially diffuse and likely spans a few tens of AU. Consequently, the obtained results represent only a mean distance to the source. 

The distance to the ribbon source can vary along the ribbon. We performed additional tests to investigate this issue. Namely, we assumed that the parallax angle changes linearly with longitude: $p=p_0+p_1(\lambda-40^\circ)$, so instead of the constant parallax angle $p$ we had two parameters $p_0$ and $p_1$. We fitted this model to the data from ESA steps 3--6, and we obtained $p_0=0.327\pm0.202$ and $p_1=-0.005\pm0.008$, with the $\chi^2_\text{min}=292.9$. Consequently, we concluded that for the selected part of the ribbon we do not have any statistically significant change in the source distance along the ribbon. However, we note that we selected only a short portion of the ribbon, so we cannot generalize this for the entire ribbon. With more data in the future one would be able to slice the ribbon for shorter segments, and perform fitting for each of them separately. This is not possible now due to limited statistics.

Inspection of the results presented in Table~\ref{tab:fitted} could suggest that the distances to the source for the higher ESA steps are closer to the observer than the lower ESA steps. However, this difference for the higher ESA steps is not statistically significant. Comparison of the $\Delta\chi^2$ estimator for different ESA steps given in Figure~\ref{fig:deltachi2} shows that they are consistent at a $1\sigma$ level with the fit to the combined data. Consequently, we can neither confirm nor reject the hypothesis that the ribbon is created at different distances depending on energy. Such a dependency is expected in the secondary ENA model \citep{zirnstein_etal:16a, zirnstein_etal:16b}, with the source distance increasing with energy. In the model by \citet{fichtner_etal:14a}, where the ribbon is created by a hydrogen wave propagating in the LISM, energy ordering results from evolution of the proton velocity distribution in the outer heliosphere \citep{sylla_fichtner:15}. Nevertheless, the uncertainties obtained for the results from individual ESA steps are too wide to identify such a subtle effect, if it exists.

The parallax baseline which spans the positions of \emph{IBEX} during observations of the same part of the sky in the ram and anti-ram hemispheres was assumed to have a constant length of 2 AU. The actual length could be a little different due to the ellipticity of Earth's orbit ($\sim$0.02 AU amplitude) and the spacecraft orbit around the Earth ($\sim$0.002 AU amplitude). Additionally, since the \emph{IBEX} spin axis only approximately follows the Sun, the spacecraft positions do not need to be exactly opposite. The potential relative bias in the parallax angle is proportional to the ratio of the effective amplitude to the 2 AU. Therefore, this effect is much smaller than the obtained uncertainty of the parallax angle.

The choice of a quadratic polynomial as the optimum curve for the ridge fitting is somewhat arbitrary.  However, we tried polynomials of various orders starting from 1, and we found that a linear function is not sufficient. For the higher-order polynomials we found that the potential improvement in the goodness of the fit does not justify using the more complex model. However, we also tried fitting to a part of the circle on the sky (a circular arc), which is also a 3-parameter model. In that case, we got an almost identical parallax angle, but the obtained values of the $\chi^2$ estimator at minimum was a little higher. Based on these findings we do not expect that the choice of a quadratic polynomial biases the final result.

The \emph{Voyagers} crossed the termination shock at distances 84 AU \citep[\emph{Voyager 2},][]{stone_etal:05a} and 94 AU \citep[\emph{Voyager 1},][]{stone_etal:08a}. The heliopause was crossed by \emph{Voyager 1} at~121 AU \citep{gurnett_etal:13a}. Our result places the ribbon source at larger distances than the heliopause found by \emph{Voyager 1}. Thus, it seems likely that the emission source of the ribbon lies in the outer heliosheath. However, the heliopause is not expected to be spherical and the distance to it in the region used in this analysis could be significantly different. 

Our analysis assumes that the \emph{IBEX}-Hi boresight is pointing exactly 90$^\circ$ off the spin axis. But an offset towards or away from the spin axis would potentially increase or decrease the observed parallax. This effect can not be distinguished from the parallax. \citet{hlond_etal:12a}, \citet{swaczyna_etal:15a} and \citet{mobius_etal:15b} discussed potential effects of the \emph{IBEX}-Lo boresight offset from the nominal position and found that the boresight deviation in the spinning direction was consistent with 0. The boresight pointing of the \emph{IBEX}-Hi detector was checked on ground to point within 0.2$^\circ$ relative to the normal to the detector baseplane \citep{funsten_etal:09a}. It is expected that \emph{IBEX}-Hi was integrated to the spacecraft with a comparable precision of 0.2$^\circ$. Additionally, the actual spin axis forms an angle of $\sim$0.1$^\circ$ with the Z-axis \citep{swaczyna_etal:15a}. Therefore, the overall uncertainty of $\sim$0.3$^\circ$ due to mounting imperfections could result from these effects. We do not have the means to assess the potential shift in inclination, but we were able to estimate it in the spinning direction. We obtained a shift of $-0.08^\circ\pm0.10^\circ$ based on data from ESA steps 3-6. This result is consistent with no deviation. 

Results of this study suggest that a higher resolution both angular and in energy, planned in the future Interstellar Mapping and Acceleration Probe (IMAP) mission \citep{mccomas_etal:11b, nrc_report:13}, would be critical from the perspective of parallax measurement. This should allow a much more accurate determination of the distance to the ribbon source. 

\section{SUMMARY AND CONCLUSIONS}
\label{sec:conclusions}

Using the data collected during the first five years of \emph{IBEX} measurements, we carefully analyzed the apparent position of the ribbon in the ram and anti-ram hemispheres in search for the geometrical parallax effect. As a baseline for our analysis, we adopted the events registered by \emph{IBEX}-Hi, histogrammed with respect to the NEP angle (spacecraft spin). After correcting the count rates for the Compton-Getting effect, they were fit to follow the Gaussian function in the proximity of the ribbon. From this we fitted the apparent position of the ribbon ridge. This position of the ridge was subsequently corrected to compensate for the shift due to frame transformation, gravity, and radiation pressure.

Based on this analysis we conclude that the observation of parallaxes of ENA sources is feasible. We present a method of data analysis with a detailed description of necessary corrections. This technique could be implemented in future missions with ENA detectors such as IMAP to better measure the distance to the ribbon source.

From comparison of the ridge apparent positions in the ram and anti-ram hemispheres, we were able to find a parallax shift of a $70^\circ$ latitude long swath of the ribbon centered at ecliptic coordinates $(\lambda,\beta)=(40^\circ,\,67^\circ)$. We did not find any statistically significant trend with the energy.  Based on the data from ESA steps 3-6, we obtained the parallax angle of $0.41^\circ\pm0.15^\circ$, which is equivalent to a distance of $140^{+84}_{-38}$~AU. These uncertainties do not include the systematic uncertainty from the mounting precision discussed in Section~\ref{sec:discussion}.

Based on the obtained parallax angles, we conclude that the distance to the ribbon source supports the hypotheses on the ribbon origin at the heliopause or in the outer heliosheath.

\acknowledgments
The authors from SRC PAS acknowledge the support by National Science Centre, Poland under grant number 2015/18/M/ST9/00036. Work by the US authors was supported by the \emph{IBEX} mission as a part of NASA's Explorer Program.


\bibliographystyle{apj}
\bibliography{iplbib,paper}

\end{document}